\documentclass[aps,prb,twocolumn,showpacs,preprintnumbers,amsmath,amssymb]{revtex4}%

\usepackage{graphicx}
\usepackage{dcolumn}
\usepackage{bm}
\usepackage{color}

\begin{document}

\title{Superconductivity at 23 K and Low Anisotropy in Rb-Substituted BaFe$_2$As$_2$ Single Crystals}

\author{Z.~Bukowski$^1$} \email{bukowski@phys.ethz.ch}
\affiliation{$^1$ Laboratory for Solid State Physics, ETH Zurich, CH-8093 Zurich, Switzerland\\
$^2$ Physik-Institut der Universit\"{a}t Z\"{u}rich, Winterthurerstrasse 190, CH-8057 Z\"urich, Switzerland\\
$^3$ Institute of Physics, Polish Academy of Sciences, Aleja Lotnik\'ow 32/46, PL-02-668 Warsaw, Poland}

\author{S.~Weyeneth$^2$} 
\affiliation{$^1$ Laboratory for Solid State Physics, ETH Zurich, CH-8093 Zurich, Switzerland\\
$^2$ Physik-Institut der Universit\"{a}t Z\"{u}rich, Winterthurerstrasse 190, CH-8057 Z\"urich, Switzerland\\
$^3$ Institute of Physics, Polish Academy of Sciences, Aleja Lotnik\'ow 32/46, PL-02-668 Warsaw, Poland}

\author{R.~Puzniak$^3$}
\affiliation{$^1$ Laboratory for Solid State Physics, ETH Zurich, CH-8093 Zurich, Switzerland\\
$^2$ Physik-Institut der Universit\"{a}t Z\"{u}rich, Winterthurerstrasse 190, CH-8057 Z\"urich, Switzerland\\
$^3$ Institute of Physics, Polish Academy of Sciences, Aleja Lotnik\'ow 32/46, PL-02-668 Warsaw, Poland}

\author{P.~Moll$^1$}
\affiliation{$^1$ Laboratory for Solid State Physics, ETH Zurich, CH-8093 Zurich, Switzerland\\
$^2$ Physik-Institut der Universit\"{a}t Z\"{u}rich, Winterthurerstrasse 190, CH-8057 Z\"urich, Switzerland\\
$^3$ Institute of Physics, Polish Academy of Sciences, Aleja Lotnik\'ow 32/46, PL-02-668 Warsaw, Poland}

\author{S.~Katrych$^1$}
\affiliation{$^1$ Laboratory for Solid State Physics, ETH Zurich, CH-8093 Zurich, Switzerland\\
$^2$ Physik-Institut der Universit\"{a}t Z\"{u}rich, Winterthurerstrasse 190, CH-8057 Z\"urich, Switzerland\\
$^3$ Institute of Physics, Polish Academy of Sciences, Aleja Lotnik\'ow 32/46, PL-02-668 Warsaw, Poland}

\author{N. D.~Zhigadlo$^1$}
\affiliation{$^1$ Laboratory for Solid State Physics, ETH Zurich, CH-8093 Zurich, Switzerland\\
$^2$ Physik-Institut der Universit\"{a}t Z\"{u}rich, Winterthurerstrasse 190, CH-8057 Z\"urich, Switzerland\\
$^3$ Institute of Physics, Polish Academy of Sciences, Aleja Lotnik\'ow 32/46, PL-02-668 Warsaw, Poland}

\author{J.~Karpinski$^1$}
\affiliation{$^1$ Laboratory for Solid State Physics, ETH Zurich, CH-8093 Zurich, Switzerland\\
$^2$ Physik-Institut der Universit\"{a}t Z\"{u}rich, Winterthurerstrasse 190, CH-8057 Z\"urich, Switzerland\\
$^3$ Institute of Physics, Polish Academy of Sciences, Aleja Lotnik\'ow 32/46, PL-02-668 Warsaw, Poland}

\author{H.~Keller$^2$}
\affiliation{$^1$ Laboratory for Solid State Physics, ETH Zurich, CH-8093 Zurich, Switzerland\\
$^2$ Physik-Institut der Universit\"{a}t Z\"{u}rich, Winterthurerstrasse 190, CH-8057 Z\"urich, Switzerland\\
$^3$ Institute of Physics, Polish Academy of Sciences, Aleja Lotnik\'ow 32/46, PL-02-668 Warsaw, Poland}

\author{B.~Batlogg$^1$}
\affiliation{$^1$ Laboratory for Solid State Physics, ETH Zurich, CH-8093 Zurich, Switzerland\\
$^2$ Physik-Institut der Universit\"{a}t Z\"{u}rich, Winterthurerstrasse 190, CH-8057 Z\"urich, Switzerland\\
$^3$ Institute of Physics, Polish Academy of Sciences, Aleja Lotnik\'ow 32/46, PL-02-668 Warsaw, Poland}

\begin{abstract}
Single crystals of Ba$_{1-x}$Rb$_{x}$Fe$_2$As$_2$ with $x=0.05-0.1$ have been grown from Sn flux and are bulk superconductors with $T_{\rm c}$ up to 23 K. The crystal structure was determined by X-ray diffraction analysis, and Sn is found to be incorporated for $\sim$ 9\% Ba, shifted by $\sim1.1$ \AA~away from the Ba site towards the (Fe$_2$As$_2$)-layers. The upper critical field deduced from resistance measurements is anisotropic with slopes of 7.1(3) T/K ($H || ab$-plane) and 4.2(2) T/K ($H || c$-axis), sufficiently far below $T_{\rm c}$. The extracted upper critical field anisotropy $\gamma_H\sim$ 3 close to $T_{\rm c}$, is in good agreement with the estimate from magnetic torque measurements. This indicates that the electronic properties in the doped BaFe$_2$As$_2$ compound are significantly more isotropic than those in the $Ln$FeAsO family. The in-plane critical current density at 5 K exceeds 10$^6$ A/cm$^2$, making Ba$_{1-x}$Rb$_x$Fe$_2$As$_2$ a promising candidate for technical applications. 

\end{abstract}

\maketitle

\section{Introduction}

The report on superconductivity at 5 K in LaFePO with the ZrCuSiAs-type structure by Kamihara \emph{et al.}\cite{one} was almost overlooked for two years until the discovery of superconductivity at $T_{\rm c}\simeq26$ K in F-substituted LaFeAsO.\cite{two} This finding initiated an intensive search for new FeAs-based superconductors and in the few following months superconductivity has been discovered in a number of analogues, primarily by substituting other rare earth ions for La, yielding its current maximum $T_{\rm c}\simeq56$ K for SmFeAsO$_{1-x}$F$_x$.\cite{three} The crystal structure of \mbox{$Ln$FeAsO} (abbreviated as ``\emph{1111}'') consists of (Fe$_2$As$_2$)-layers sandwiched by ($Ln_2$O$_2$)-layers, where $Ln$ denotes any lanthanide element. The parent compound \mbox{$Ln$FeAsO} is antiferromagnetic, but may become superconducting upon electron doping by either partially replacing oxygen by fluorine, by generating oxygen deficiency, or by applying pressure.\cite{four} Superconductivity can be also induced in $Ln$FeAsO through electron doping (partially replacing $Ln$ by Th\cite{five1}) or hole doping (partially replacing $Ln$ by Sr\cite{six1}).

\begin{figure}[b!]
\includegraphics[width=1\linewidth]{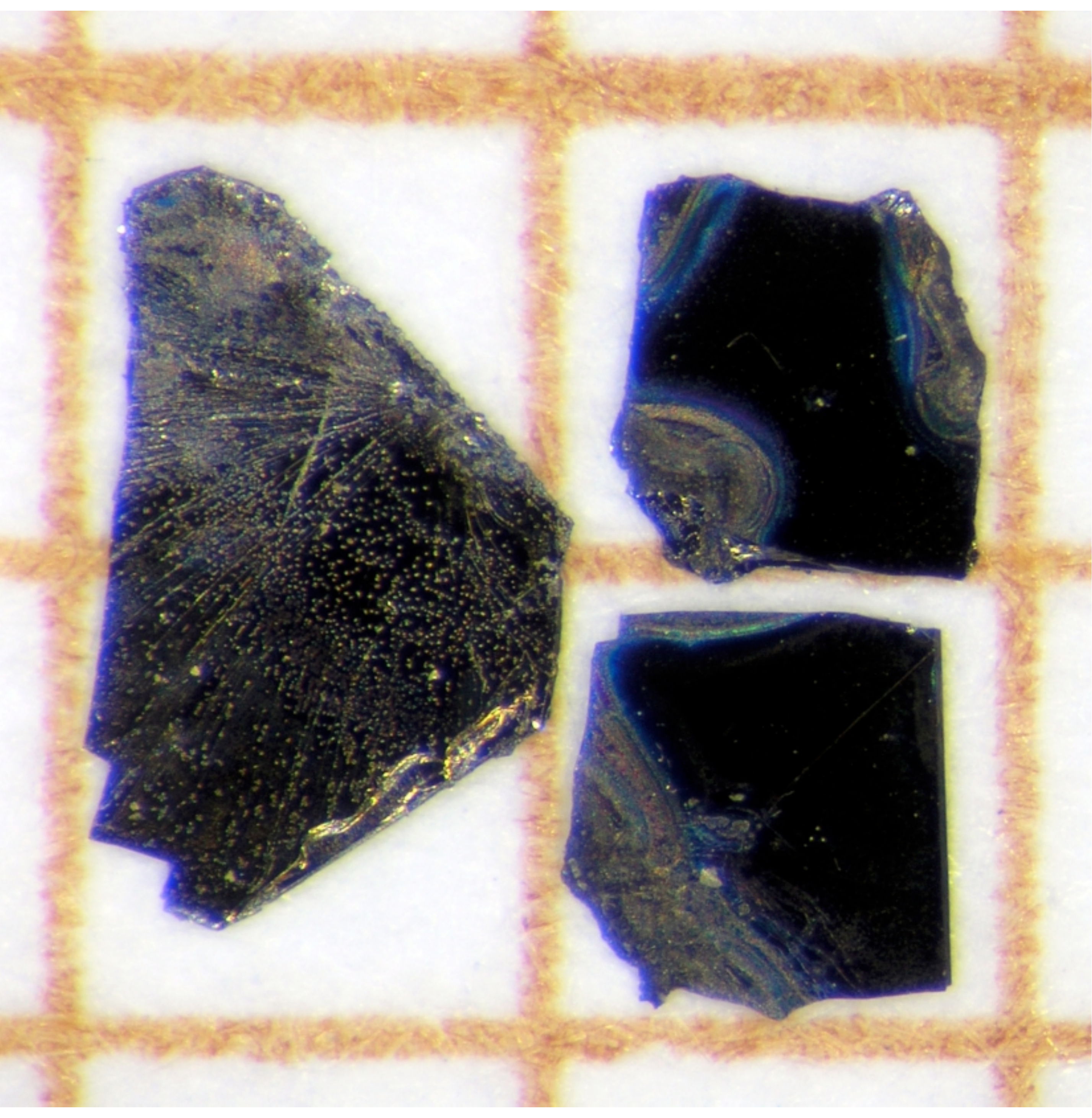}
\vspace{0cm}
\caption{(Color online) Photograph of three single crystals of Ba$_{0.84}$Rb$_{0.10}$Sn$_{0.09}$Fe$_2$As$_{1.96}$ on a millimeter grid.}
\label{fig1}
\end{figure}

More recently, the discovery of superconductivity at 38 K in Ba$_{1-x}$K$_x$Fe$_2$As$_2$ with the ThCr$_2$Si$_2$-type structure has been reported.\cite{seven1} The $A$Fe$_2$As$_2$ ($A$ = Ca, Sr, Ba) compounds (called ``\emph{122}'') have a more simple crystal structure in which (Fe$_2$As$_2$)-layers, identical to those in $Ln$FeAsO, are separated by single elemental $A$ layers. Thus, a new class of superconductors was established by the subsequent reports on superconductivity in isostructural hole-doped Sr$_{1-x}$K$_x$Fe$_2$As$_2$ and Sr$_{1-x}$Cs$_x$Fe$_2$As$_2$,\cite{eight1} Ca$_{1-x}$Na$_x$Fe$_2$As$_2$,\cite{nine1} Eu$_{1-x}$K$_x$Fe$_2$As$_2$ and Eu$_{1-x}$Na$_x$Fe$_2$As$_2$,\cite{ten1,eleven1} and in electron-doped Co-substituted BaFe$_2$As$_2$ and SrFe$_2$As$_2$,\cite{twelve1,thirteen1} and Ni-substituted BaFe$_2$As$_2$.\cite{fourteen1} Furthermore, pressure induced superconductivity has been discovered in the parent compounds CaFe$_2$As$_2$,\cite{fifteen1, sixteen1} SrFe$_2$As$_2$,\cite{seventeen1, eighteen1} and BaFe$_2$As$_2$.\cite{eighteen1}

Besides KFe$_2$As$_2$ and CsFe$_2$As$_2$, which are superconductors with $T_{\rm c}$'s of 3.8 K and 2.6 K,\cite{eight1} respectively, RbFe$_2$As$_2$ is known to exist as well.\cite{nineteen1} Therefore, it seemed natural to us using Rb as a chemical substituent in order to extend the number of elements which can effectively induce superconductivity in $A$Fe$_2$As$_2$ compounds. In this paper, we report on the superconductivity induced in BaFe$_2$As$_2$ by partial substitution of Rb for Ba, and present its basic superconducting properties, including estimates of the electronic anisotropy.

\begin{figure}[b!]
\includegraphics[width=1\linewidth]{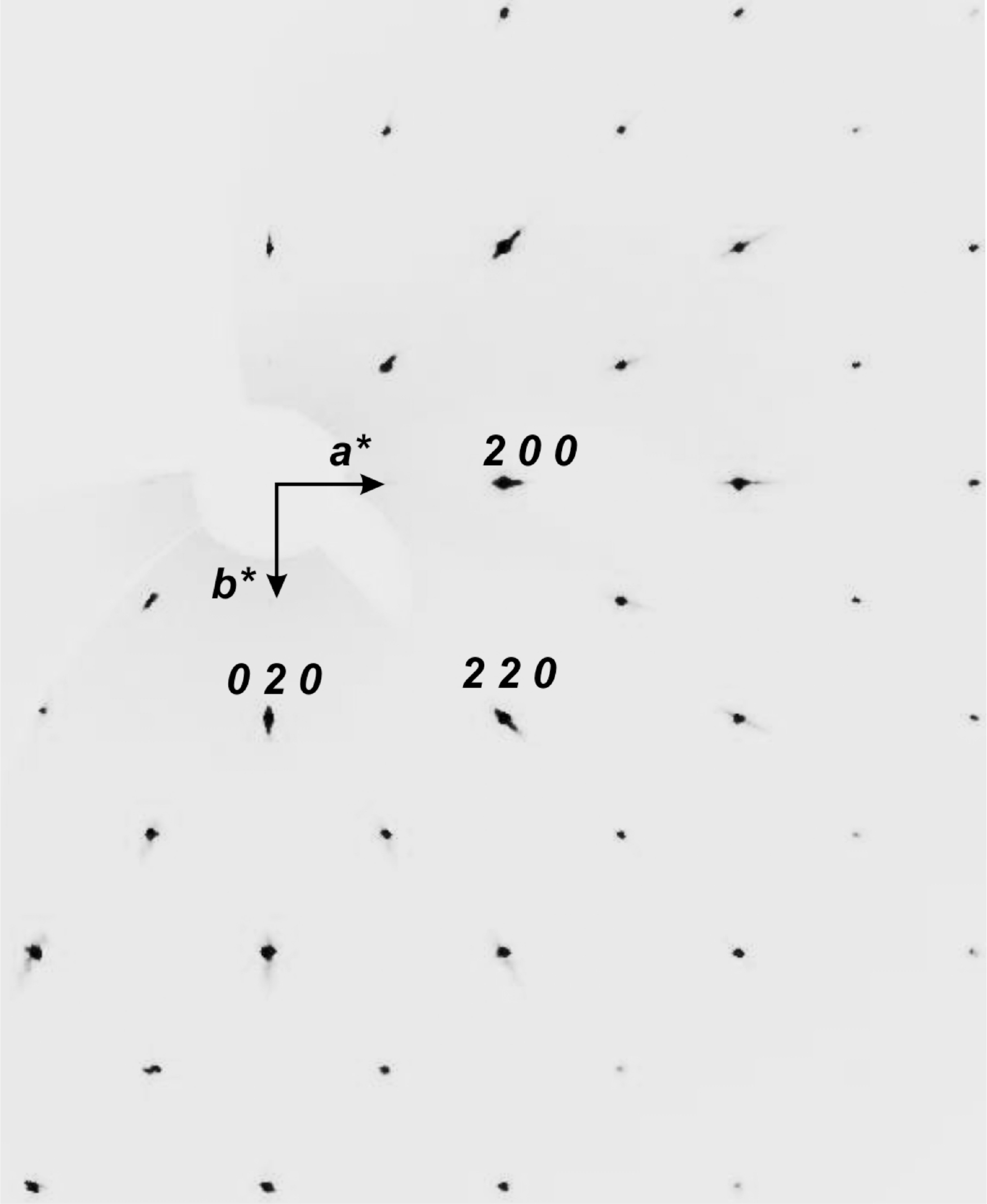}
\vspace{0cm}
\caption{The \emph{hk0} reciprocal space section determined by XRD of the single crystal Ba$_{0.84}$Rb$_{0.10}$Sn$_{0.09}$Fe$_2$As$_{1.96}$.}
\label{fig2}
\end{figure}

\section{Experimental details}
Single crystals of Rb substituted BaFe$_2$As$_2$ [(Ba,Rb)122] were grown using a Sn flux method similar to that described by Ni \emph{et al.}\cite{twenty1} The Fe:Sn ratio (1:24) in a starting composition was kept constant in all runs while the Rb:Ba ratio was varied between 0.7 and 2.0. The appropriate amounts of Ba, Rb, Fe$_2$As, As, and Sn were placed in alumina crucibles and sealed in silica tubes under 1/3 atmosphere of Ar gas. Next, the ampoules were heated at 850 $^\circ$C for 3 hours until all components were completely melted, and cooled to 500 $^\circ$C in 50 hours. At this temperature the ampoule was turned upside down inside a furnace and liquid Sn was decanted from the crystals. The remaining thin film of Sn at the crystal surfaces was subsequently dissolved at room temperature by soaking crystals for a few days in liquid Hg. Finally, the crystals were heated for one hour at 190 $^\circ$C in vacuum to evaporate the remaining traces of Hg. No signs of superconducting Hg are seen in the magnetic measurements.

The phase purity was checked on crushed crystals by means of powder X-ray diffraction (XRD) measurements carried out on a STOE diffractometer using Cu-K$_\alpha$ radiation and a graphite monochromator.

\begin{figure}[t!]
\includegraphics[width=1\linewidth]{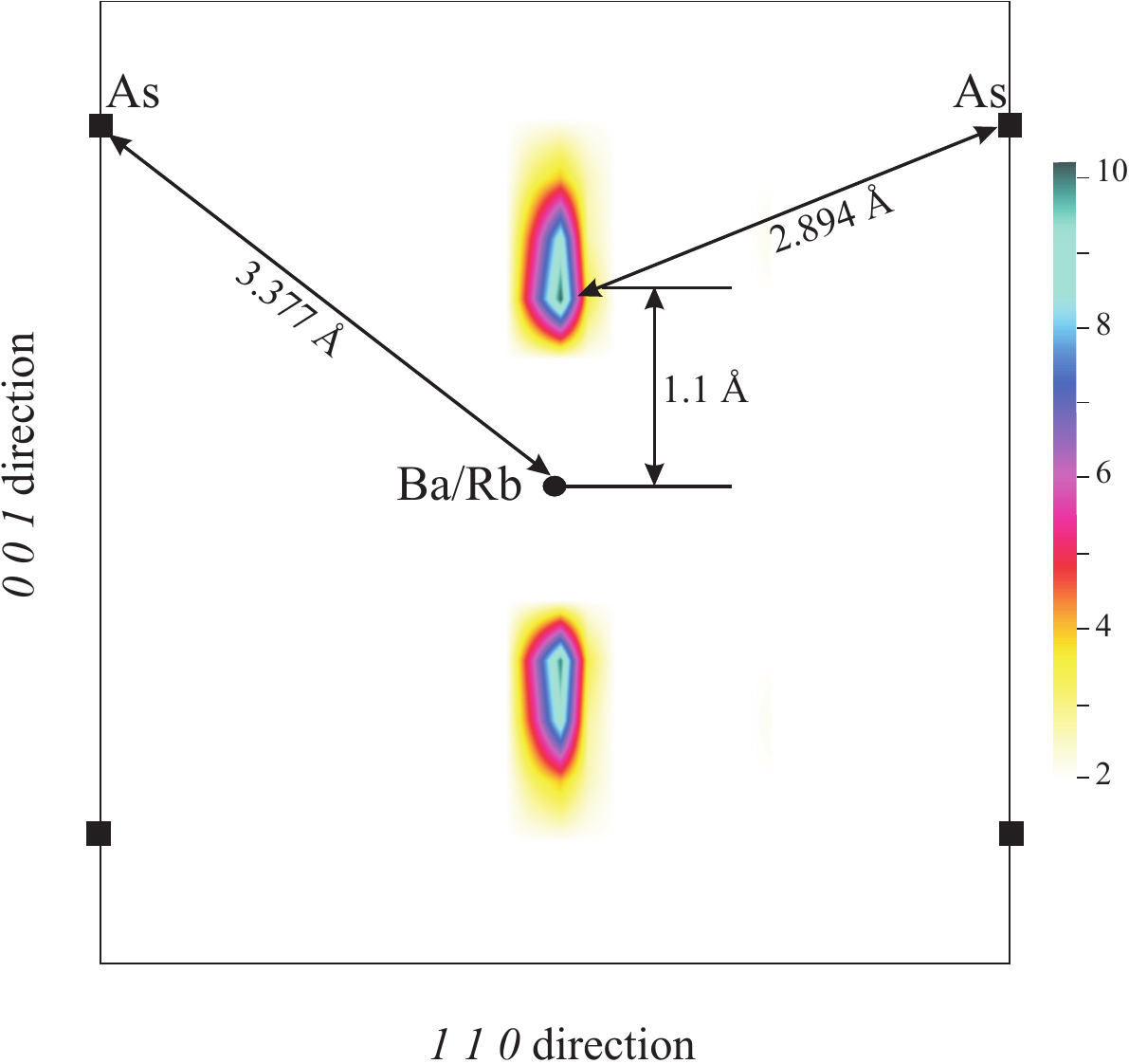}
\vspace{0cm}
\caption{(Color online) \emph{110} section of the $F_0-F_c$ difference Fourier map. The enhanced electron density reveals the location of Sn atoms, shifted by 1.1 \AA~towards the (Fe$_2$As$_2$)-layers. The black dot shows the Ba/Rb position and black squares mark the As positions.}
\label{fig3}
\end{figure}

Single-crystal X-ray diffraction data were collected on a 4-circle diffractometer equipped with a CCD detector (Oxford Diffraction Ltd, Mo-K$_\alpha$, 60 mm sample to detector distance). Data reduction and analytical absorption correction were applied using the CrysAlis RED software package.\cite{twentyone1} The crystal structure was refined on $F^2$ employing the SHELXL program.\cite{twentytwo1} The starting model for the refinement was taken from Ref. \cite{twentythree1} The elemental analysis of the crystals was performed by means of energy dispersive X-ray (EDX) spectrometry.

Magnetic measurements were performed in a Quantum Design Magnetic Property Measurement System (MPMS XL) with the Reciprocating Sample Option (RSO) installed. Transport measurements were performed in a 14 Tesla Quantum Design Physical Property Measurement System (PPMS). Magnetic torque measurements were carried out by using a highly sensitive miniaturized piezoresistive torque sensor within a home made experimental set-up described elsewhere.\cite{twentyfour1,twentyfive1}

\begin{figure}[b!]
\includegraphics[width=1\linewidth]{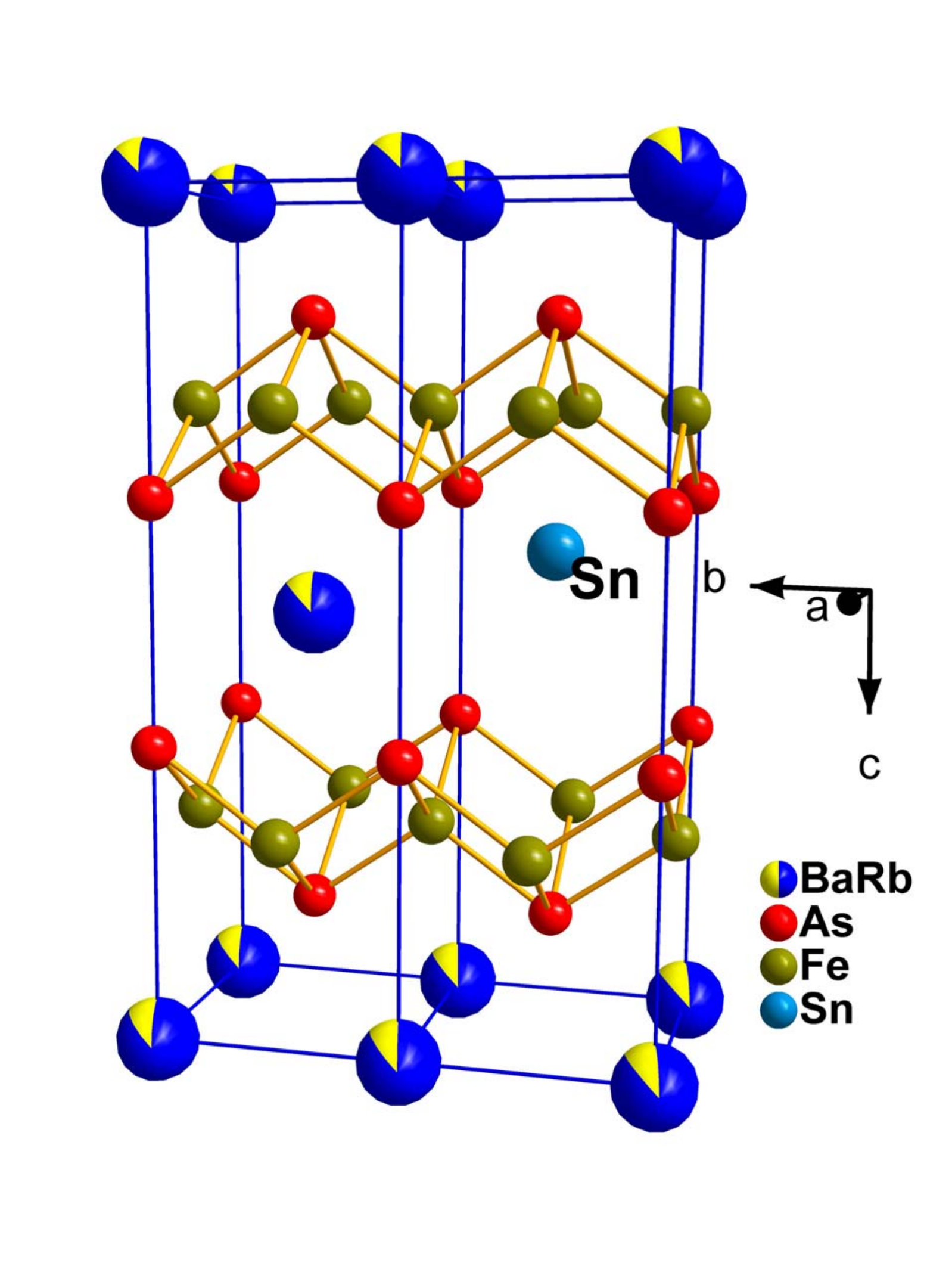}
\vspace{0cm}
\caption{(Color online) Schematic illustration of two unit cells of Rb and Sn substituted BaFe$_2$As$_2$. The possible Sn location is shown, for clarity, on one site only.}
\label{fig4}
\end{figure}

\section{Results and discussion}
The single crystals of Ba$_{1-x}$Rb$_x$Fe$_2$As$_2$ grow in a plate-like shape with typical dimensions (1-3) x (1-2) x (0.05-0.1) mm$^3$ (see Fig. \ref{fig1}). Depending on the starting composition, the crystals displayed a broad variety of properties from non-superconducting to superconducting with rather sharp transitions to the superconducting state. For further studies we chose single crystals grown from the starting composition Ba$_{0.6}$Rb$_{0.8}$Fe$_2$As$_2$. The composition of the crystals from this batch determined by EDX analysis (16.79 at.\% Ba, 1.94 at.\% Rb, 1.74 at.\% Sn, 40.19 at.\% Fe, and 39.33 at.\% As) leads to the chemical formula Ba$_{0.84}$Rb$_{0.10}$Sn$_{0.09}$Fe$_2$As$_{1.96}$. Crystals from the selected batch exhibit a $T_{\rm c}$ around 23 K but compared to the crystals with higher $T_{\rm c}$ their superconducting transition is relatively sharp, suggesting superior quality.

\subsection{Crystal structure}
The crystals studied by XRD are of good quality, and no additional phases (impurities, twins or intergrowing crystals) were detected by examining the reconstructed reciprocal space sections (see Fig. \ref{fig2}). The average mosaic spread of 1.45$^\circ$ was estimated using the XCalibur, CrysAlis Software System by analyzing all frames.\cite{twentyone1}

\begin{table}[b!]
\caption{Crystal data and structure refinement for Rb and Sn substituted BaFe$_2$As$_2$.}
\begin{center}
\begin{tabular}{| l | l |}
\hline
Crystallographic formula (XRD) 	& Ba$_{0.89}$Rb$_{0.05}$Sn$_{0.06}$Fe$_2$As$_{2}$ 	\\\hline
Temperature (K) 				& 295(2)										\\\hline
Wavelength (\AA) 				& 0.71073/Mo-K$_\alpha$  						\\\hline
Crystal system, space group, Z 	& Tetragonal, $I4$/mmm, 2   						\\\hline
Unit cell dimensions (\AA)			& $a$= 3.9250(2), $c$=13.2096(5)					\\\hline
Volume (\AA$^3$)				& 203.502(3)									\\\hline
Calculated density (g/cm$^3$)		& 6.449										\\\hline
Absorption correction type		& analytical									\\\hline
Absorption coefficient (mm$^{-1}$)	& 32.413										\\\hline
F(000)						& 345										\\\hline
Crystal size ($\mu$m$^3$) 		& 117 x 77 x 18									\\\hline
$\theta$ range for data collection	& 5.42 to 42.81 deg								\\\hline
Index ranges					& $-6\leqslant h\leqslant7$, $-7\leqslant k\leqslant6$,	\\ 	
							& $-26\leqslant l\leqslant7$						\\\hline
Reflections collected/unique		& 1626/291 $R_{\rm int}=0.0458$						\\\hline
Completeness to 2 $\theta$		& 96.4 \%										\\\hline
Refinement method				& Full-matrix least-squares						\\
							& on $F^2$									\\\hline
Data/restraints/parameters		& 291/0/13									\\\hline
Goodness-of-fit on $F^2$			& 1.052										\\\hline
Final $R$ indices [$I>2\sigma(I)$]	& $R_1$= 0.0389, $wR_2$ = 0.1106				\\\hline
$R$ indices (all data)			& $R_1$=0.0414, $wR_2$= 0.1122					\\\hline
$\Delta\rho_{\rm max}$ and $\Delta\rho_{\rm min}$ (e/\AA$^3$)		& 5.848 and -3.457			\\\hline
Bond lengths (\AA)				& 											\\\hline
Ba/Rb-As						& 3.3774(3) x 8									\\\hline
Fe-As						& 2.3979(3) x 4									\\\hline
Fe-Fe						& 2.7754(1) x 4									\\\hline
As-Sn						& 2.894(3) x 4									\\\hline
Fe-Sn						& 2.945(7) x 4									\\\hline
Bond angles (deg)				& 											\\\hline
As-Fe-As						& 109.86(2)									\\
							& 109.28(1)									\\\hline

\end{tabular}
\end{center}
\label{tab1}
\end{table}

We assumed that Rb atoms substitute for Ba atoms and the Rb/Ba occupations have been refined simultaneously. The content of these elements was found to be 1.0 at.\% (Rb) and 17.8 at.\% (Ba), in acceptable agreement with the EDX analysis (1.9 at.\% Rb and 16.8 at.\% Ba).

It has been reported that the BaFe$_2$As$_2$ crystals grown from a Sn flux have approximately 1 at.\% of Sn incorporated into the structure with the Sn atoms most likely located on the As sites.\cite{twenty1} However, our structure refinement reveals a somewhat different picture. After several cycles of refinement the Fourier difference map shows two pronounced maxima of the electron density away from the Ba/Rb site. We located the Sn atoms on these sites, shifted towards the (Fe$_2$As$_2$)-layers (Fig. \ref{fig3}). This interpretation is supported by the considerable reduction of the $R$ factor from 5.41 \% to 3.89 \% when Sn atoms are allowed to occupy these ``off-center'' sites. The resulting distances to the As site reflect very well the size difference between Ba and Sn, considering the covalent radius. 


\begin{table}[t!]
\caption{Atomic coordinates, occupancy factors, and equivalent isotropic and anisotropic displacement parameters [\AA$^2$ x $10^3$] for Rb and Sn substituted BaFe$_2$As$_2$.}
\begin{center}
\begin{tabular}{|c|c|c|c|c|c|c|c|c|}
\hline
Atom	 & Site & $x$ & $y$ & $z$ 	& $Occ.$ & $U_{\rm iso}$ & $U_{11}=$ & $U_{33}$		 		\\
		&		&		&		&		&		&		& $U_{22}$	&	\\\hline
Ba(Rb)	& $2a$	& 0		& 0		& 0	 	& 0.89(0.05) & 17(1)	& 17(1) 	& 17(1)  	\\
Sn		& $4e$	& 0		& 0		& 0.0837(7) &  0.06	& 10(3)	& 13(4) 	& 5(3)  	\\
As		& $4e$	& 0		& 0		& 0.3543(1) & 1 	& 12(1)	& 11(1) 	& 15(1)  	\\
Fe		& $4d$	& 0.5 	& 0		& 0.25 	& 1	 	& 12(1)	& 9(1) 	& 17(1)  	\\\hline
\end{tabular}
\end{center}
\emph{$U_{\rm iso}$ is defined as one third of the trace of the orthogonalized $U_{ij}$ tensor. The anisotropic displacement factor exponent takes the form: $-2\pi^2\cdot(h^2a^2\cdot U_{11} + ... + 2hkab\cdot U_{12})$. For symmetry reasons $U_{23}=U_{13}=U_{12}=0.$}
\end{table}


\begin{figure}[b!]
\includegraphics[width=1\linewidth]{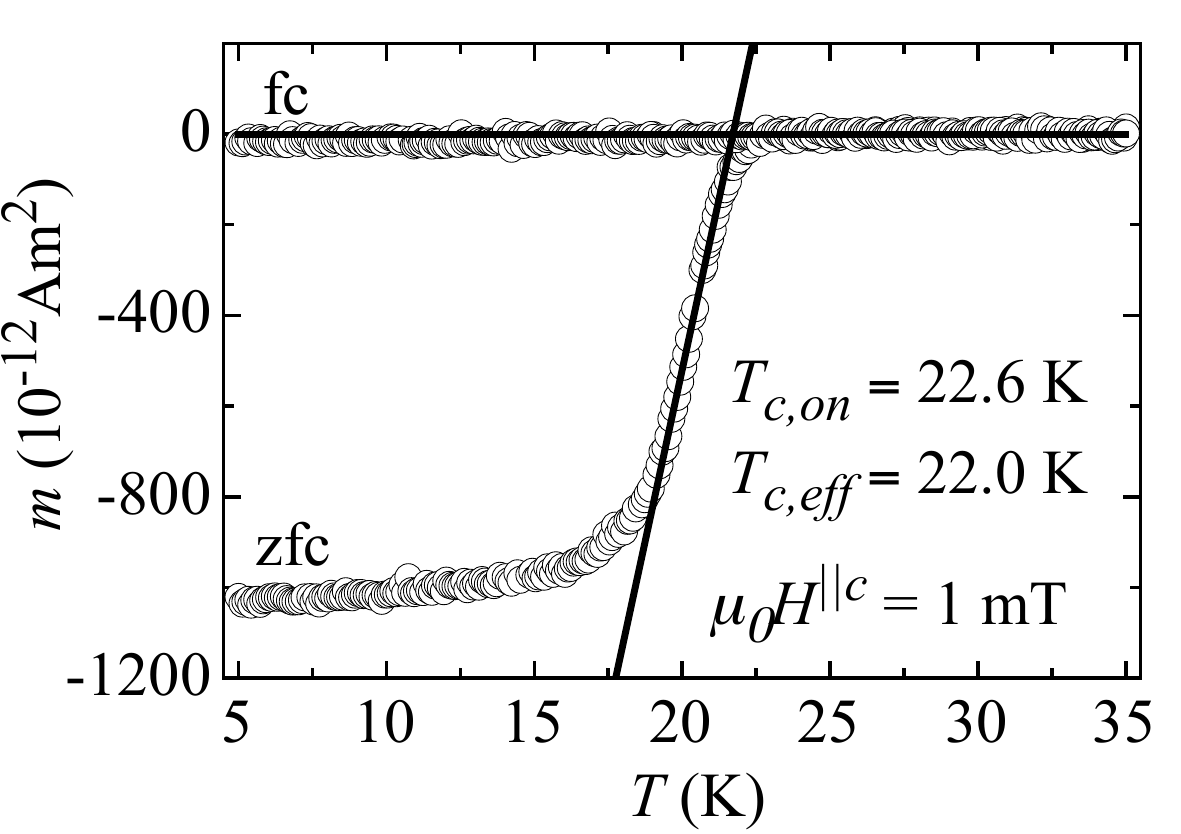}
\vspace{0cm}
\caption{Temperature dependence (field-cooled and zero-field-cooled) of the magnetic moment in a magnetic field of 1 mT applied parallel to the $c$-axis on the Ba$_{0.84}$Rb$_{0.10}$Sn$_{0.09}$Fe$_2$As$_{1.96}$ crystal with a volume of about $1.56\cdot10^{-13}$ m$^3$.}
\label{fig5}
\end{figure}

We assumed the overall occupation of Ba, Rb, and Sn to be 100 \%. The Sn occupation was also refined and found to be $\sim$ 6 \% of the Ba/Rb/Sn sites. The Sn content of $1.2\pm0.3$ at.\% agrees with the EDX data (1.7 at.\%). The results of structure refinement are presented in the Tables 1 and 2. The resulting structure is shown in Fig. \ref{fig4}. Compared to unsubstituted BaFe$_2$As$_2$ the lattice parameter $a$ is slightly shorter, the $c$ parameter is longer and the volume of the unit cell is smaller too.\cite{twentythree1} A similar tendency has been observed for other ``\emph{122}'' compounds, when Ba or Sr is replaced by K.\cite{eight1, twentysix1, twentyseven1} The increase of the $c$ parameter in the studied crystals Ba$_{0.84}$Rb$_{0.10}$Sn$_{0.09}$Fe$_2$As$_{1.96}$ is caused mainly by substitution of Ba$^{2+}$ ions ($r=1.42$ \AA) by larger Rb$^+$ ions ($r=1.61$ \AA).\cite{twentyeight1} The relatively marked shortening of the $a$ parameter (larger than expected from Vegard$^{\prime}$s law) seems to be caused by Sn incorporation.

\subsection{Critical current density and irreversibility line}

A plate like single crystal from the same batch and therefore identical lattice parameters with approximate dimensions of 125 x 125 x 10 $\mu$m$^3$ was chosen for DC magnetization and for magnetic torque studies. In Fig. \ref{fig5} we show the susceptibility measured in a magnetic field of 1 mT parallel to the crystallographic $c$-axis, showing a narrow transition with an effective transition temperature around 22 K and with an onset to superconductivity at 22.6 K.

\begin{figure}[t!]
\includegraphics[width=1\linewidth]{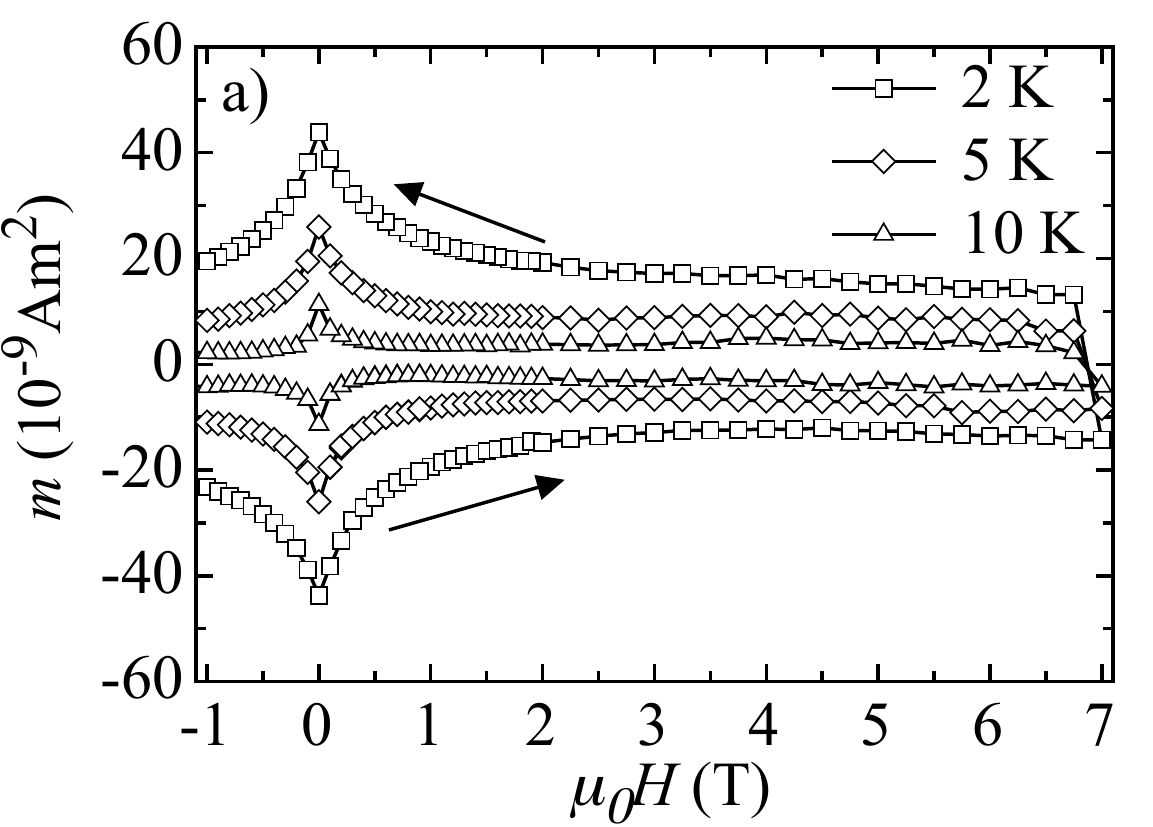}
\includegraphics[width=1\linewidth]{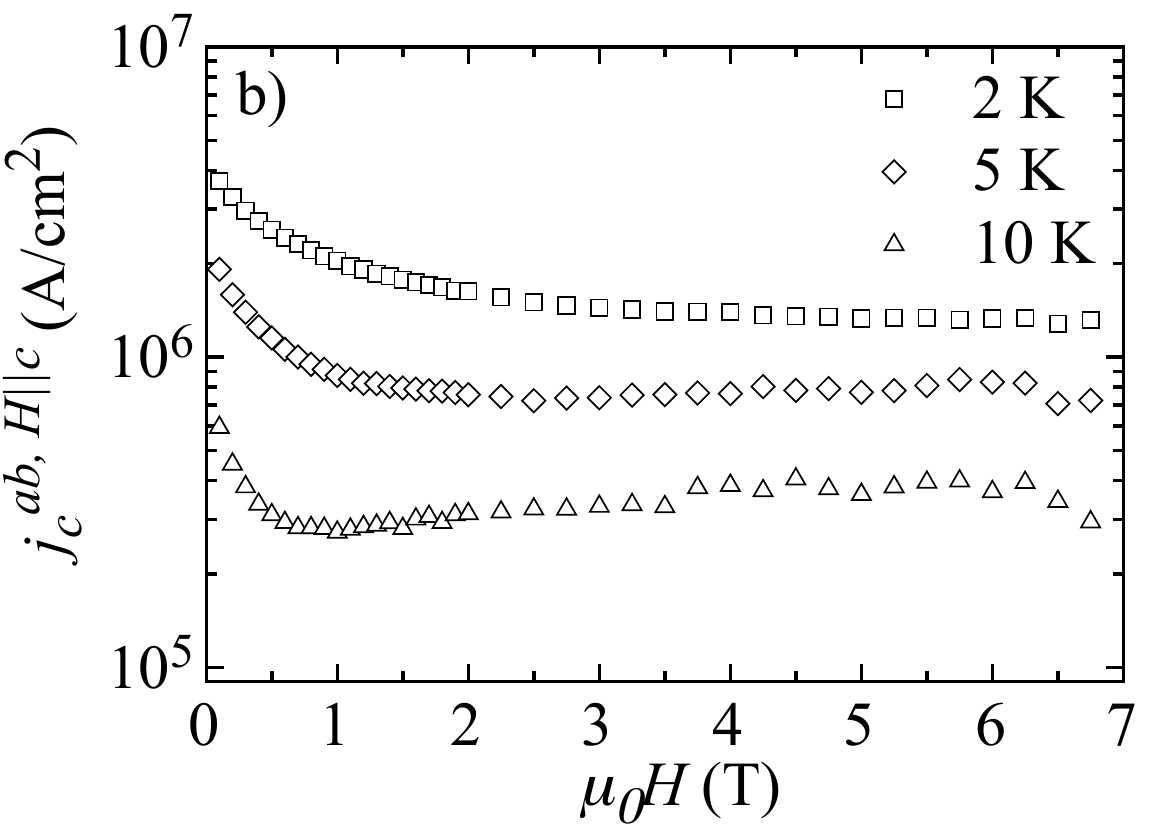}
\vspace{0cm}
\caption{a) Magnetic hysteresis loops measured at 2 K, 5 K, and 10 K in a field up to 7 T parallel to the $c$-axis on the Ba$_{0.84}$Rb$_{0.10}$Sn$_{0.09}$Fe$_2$As$_{1.96}$ crystal with a volume of about $1.56\cdot10^{-13}$ m$^3$ and with dimensions of $125$ x $125$ $\mu$m$^2$ in the plane perpendicular to the applied magnetic field. b) The critical current density calculated from the hysteresis loops.}
\label{fig6}
\end{figure}

The low temperature signal recorded in the zero-field-cooled mode corresponds to a full diamagnetic response. The observed extremely small magnetic moment in the field-cooled mode is due to the pronounced magnetic irreversibility, possibly due to the local lattice distortions caused by the substitution with relatively big Rb ions, introducing effective pinning centers. 

\begin{figure}[t!]
\includegraphics[width=1\linewidth]{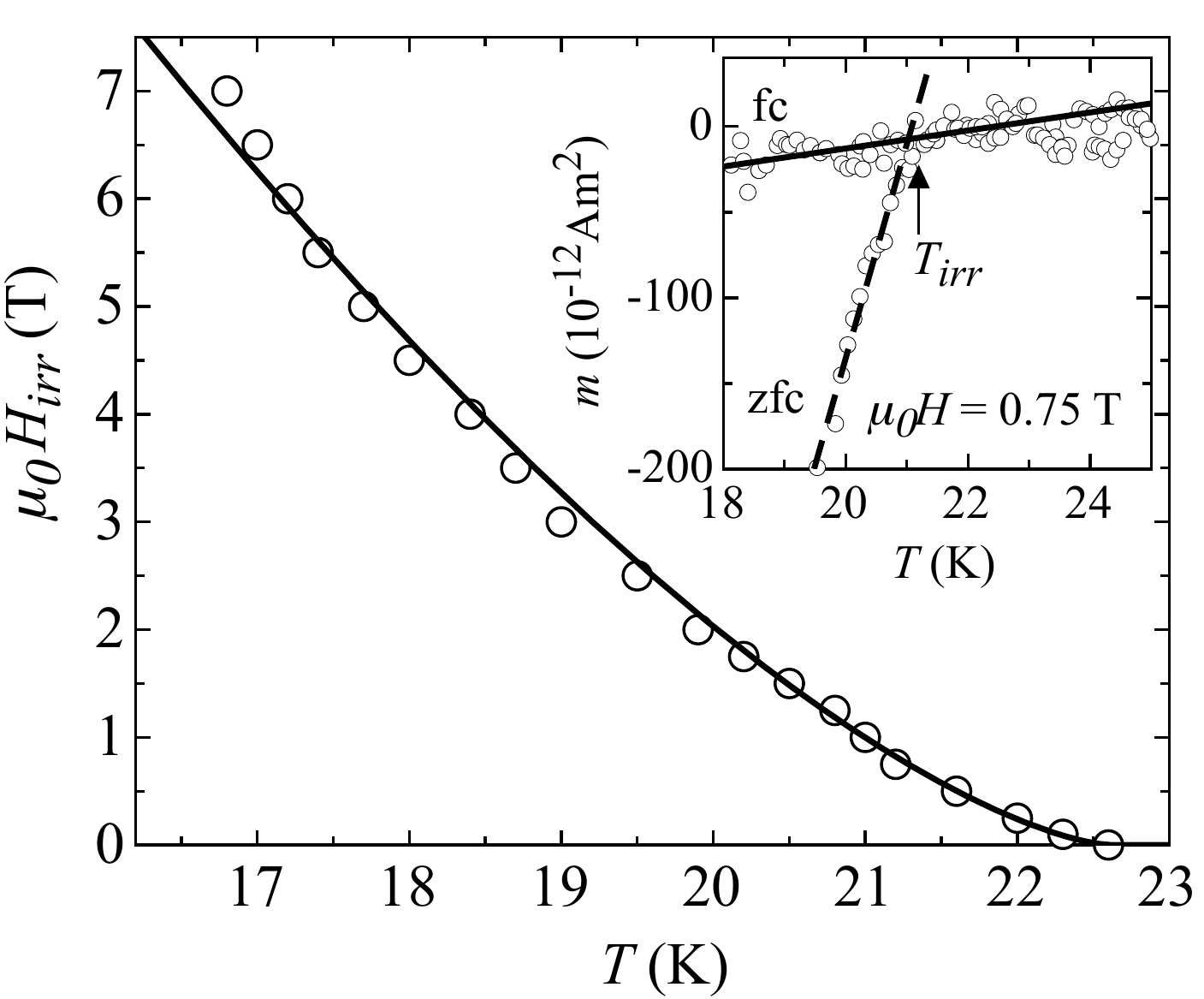}
\vspace{0cm}
\caption{Temperature dependence of the irreversibility field $H_{\rm irr}$ for Ba$_{0.84}$Rb$_{0.10}$Sn$_{0.09}$Fe$_2$As$_{1.96}$ for $H||c$. The inset illustrates how the quantity was determined. The irreversibility line is approximated very well by a power-law temperature dependence with an exponent 3/2 (solid line), as discussed in the text.}
\label{fig7}
\end{figure}

A relatively strong pinning was confirmed in magnetic hysteresis loop measurements (see Figs. \ref{fig6}a, \ref{fig6}b, and \ref{fig7}) and by magnetic torque, as discussed later. The critical current density at 2 K, 5 K, and 10 K, estimated from the field dependence of the magnetic moment using Bean's model,\cite{twentynine1, thirty1} reaches values of the order of $10^6$ A/cm$^2$ (see Fig. \ref{fig6}b), which is very promising for applications. Similar values for the critical current were reported for BaFe$_2$As$_2$ substituted with K.\cite{thirtyone1} The slight increase of the critical current density with increasing field in Fig. \ref{fig6}b is most likely due to the peak effect, which results in an effective increase of the irreversibility in the $M(H)$ curves (see Fig. \ref{fig6}a). Numerous explanations have been proposed, relying the effect to an increase in the microscopic pinning force, matching effects, field induced granularity or pinning site activation, crossover of pinning regimes, or a phase transition in vortex matter. All models include a field dependent flux-creep rate and a critical current density that decreases monotonically with increasing magnetic field. \cite{thirtytwo}

From temperature dependent magnetization measurements at various magnetic fields we deduced the irreversibility line $H_{\rm irr}(T)$ by following the temperatures for which the zero-field-cooled and field-cooled branches merge. The results are plotted in Fig. \ref{fig7}, where the upper inset to the figure illustrates the derivation in a magnetic field of 0.75 T. The irreversibility line is located in relatively high magnetic fields. A similar behavior has been reported for K substituted BaFe$_2$As$_2$.\cite{thirtyone1}

\begin{figure}[b!]
\vspace{0cm}
\includegraphics[width=1\linewidth]{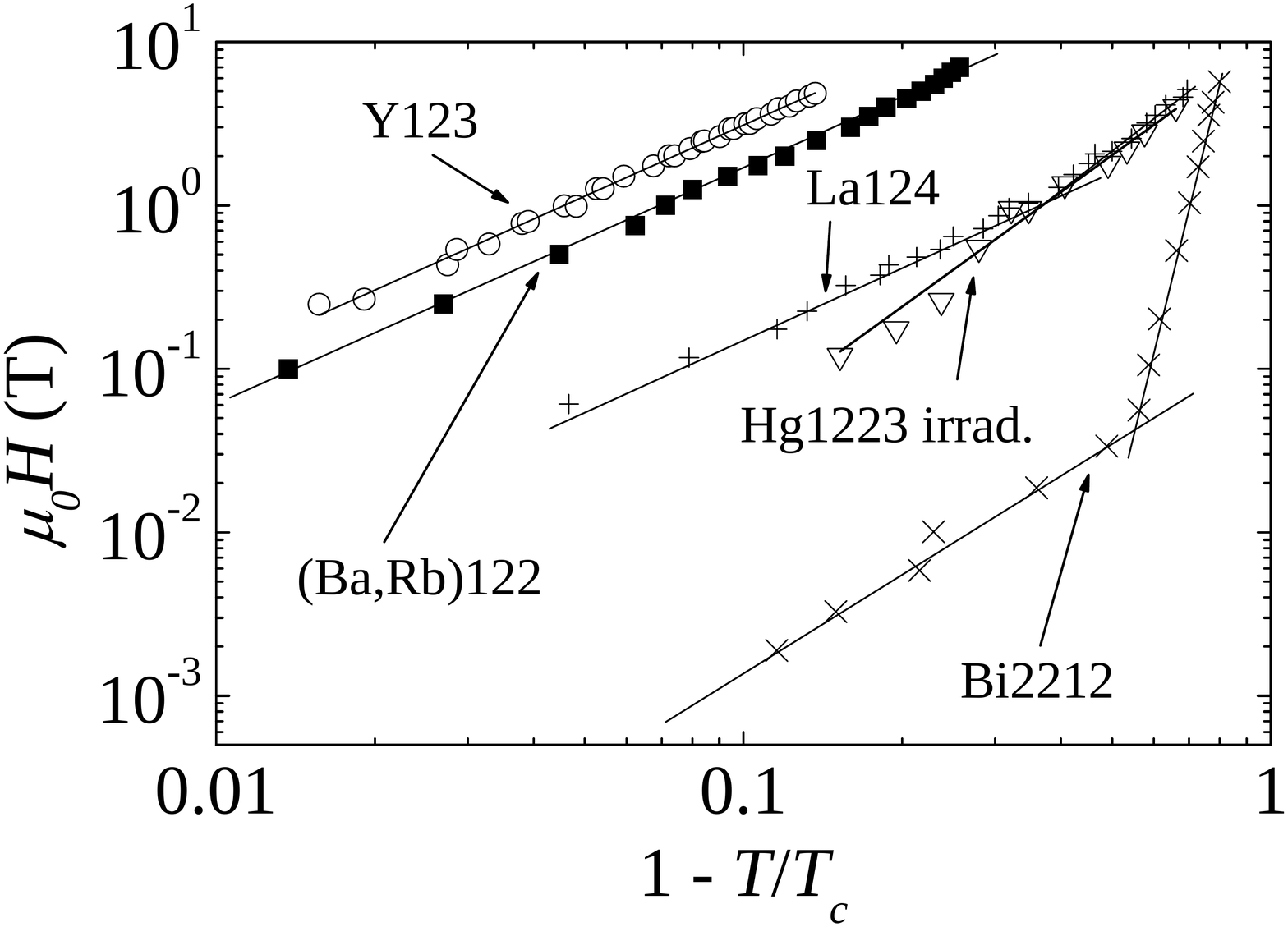}
\vspace{0cm}
\caption{Comparison of the irreversibility line for the Ba$_{0.84}$Rb$_{0.10}$Sn$_{0.09}$Fe$_2$As$_{1.96}$ single crystal for $H||c$-axis [(Ba,Rb)122] with those for YBa$_2$Cu$_3$O$_{7-\delta}$ [Y123],\cite{thirtyfour} La$_{1.86}$Sr$_{0.14}$CuO$_4$ [La214],\cite{thirtyfive} Bi$_2$Sr$_2$CaCu$_2$O$_{8+\delta}$ single crystals [Bi2212]\cite{thirtysix} and with HgBa$_2$Ca$_2$Cu$_3$O$_{8+\delta}$ single crystal with strong pinning centers intentionally introduced by neutron irradiation [Hg1223 irrad.].\cite{thirtyseven} The solid lines are fits to a power-law dependence as explained in the text.}
\label{fig8}
\end{figure}

The irreversibility line for Ba$_{0.84}$Rb$_{0.10}$Sn$_{0.09}$Fe$_2$As$_{1.96}$ is very well described by a power-law temperature dependence according to $(1 - T/T_{\rm c})^n$ with fitted parameters $T_{\rm c}=22.6(2)$K and $n=1.47(5)$ (see the straight line in the log-log $H_{\rm irr}$ vs. $(1- T/T_{\rm c})$ dependence in Fig. \ref{fig8}). The value of $n=1.47(5)$ is very close to $n=3/2$, typical for high-$T_{\rm c}$ superconductors characterized by an unusually small coherence length and an exceptionally high thermal activation at high temperatures.\cite{thirtythree} A comparison of the irreversibility lines is presented in Fig. \ref{fig8}: single crystal of Ba$_{0.84}$Rb$_{0.10}$Sn$_{0.09}$Fe$_2$As$_{1.96}$, high-$T_{\rm c}$ YBa$_2$Cu$_3$O$_{7-\delta}$,\cite{thirtyfour} La$_{1.86}$Sr$_{0.14}$CuO$_4$,\cite{thirtyfive} Bi$_2$Sr$_2$CaCu$_2$O$_{8+\delta}$ single crystals,\cite{thirtysix} and HgBa$_2$Ca$_2$Cu$_3$O$_{8+\delta}$ single crystal with strong pinning centers intentionally introduced by neutron irradiation.\cite{thirtyseven} The irreversibility line for Ba$_{0.84}$Rb$_{0.10}$Sn$_{0.09}$Fe$_2$As$_{1.96}$ is located in significantly higher magnetic fields and temperatures than those of La$_{1.86}$Sr$_{0.14}$CuO$_4$, Bi$_2$Sr$_2$CaCu$_2$O$_{8+\delta}$, and HgBa$_2$Ca$_2$Cu$_3$O$_{8+\delta}$. Its position in the reduced temperature phase diagram is comparable only with that one for YBa$_2$Cu$_3$O$_{7-\delta}$, i.e., with the position of the irreversibility line for the high-$T_{\rm c}$ superconductor characterized by the lowest anisotropy value among the compounds compared here ($\gamma\sim6$). Furthermore, the power-law exponents describing the irreversibility lines for both Ba$_{0.84}$Rb$_{0.10}$Sn$_{0.09}$Fe$_2$As$_{1.96}$ and YBa$_2$Cu$_3$O$_{7-\delta}$ are essentially identical (see the parallel lines in Fig. \ref{fig8}).

\begin{figure}[t!]
\vspace{0cm}
\includegraphics[width=1\linewidth]{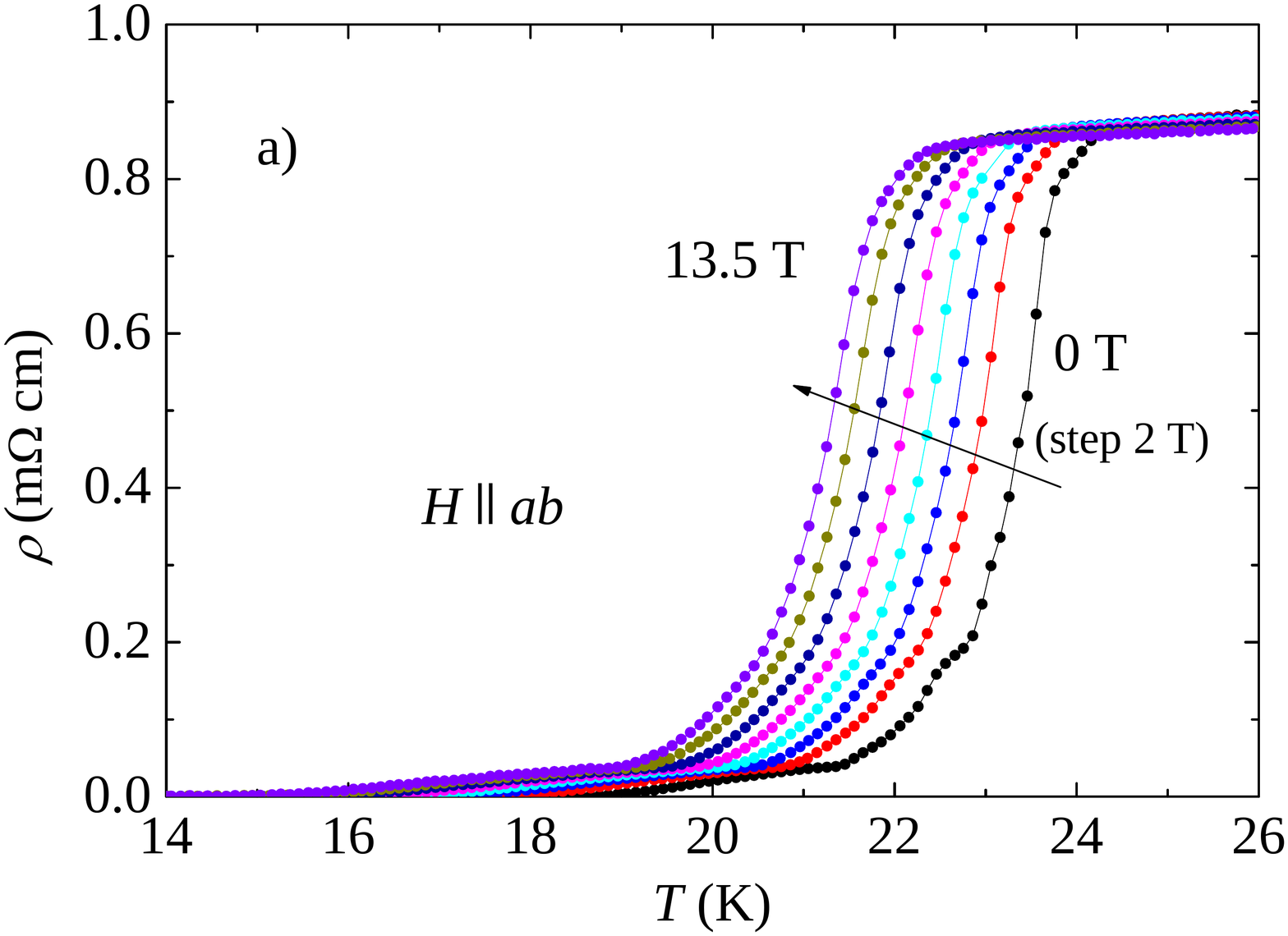}
\includegraphics[width=1\linewidth]{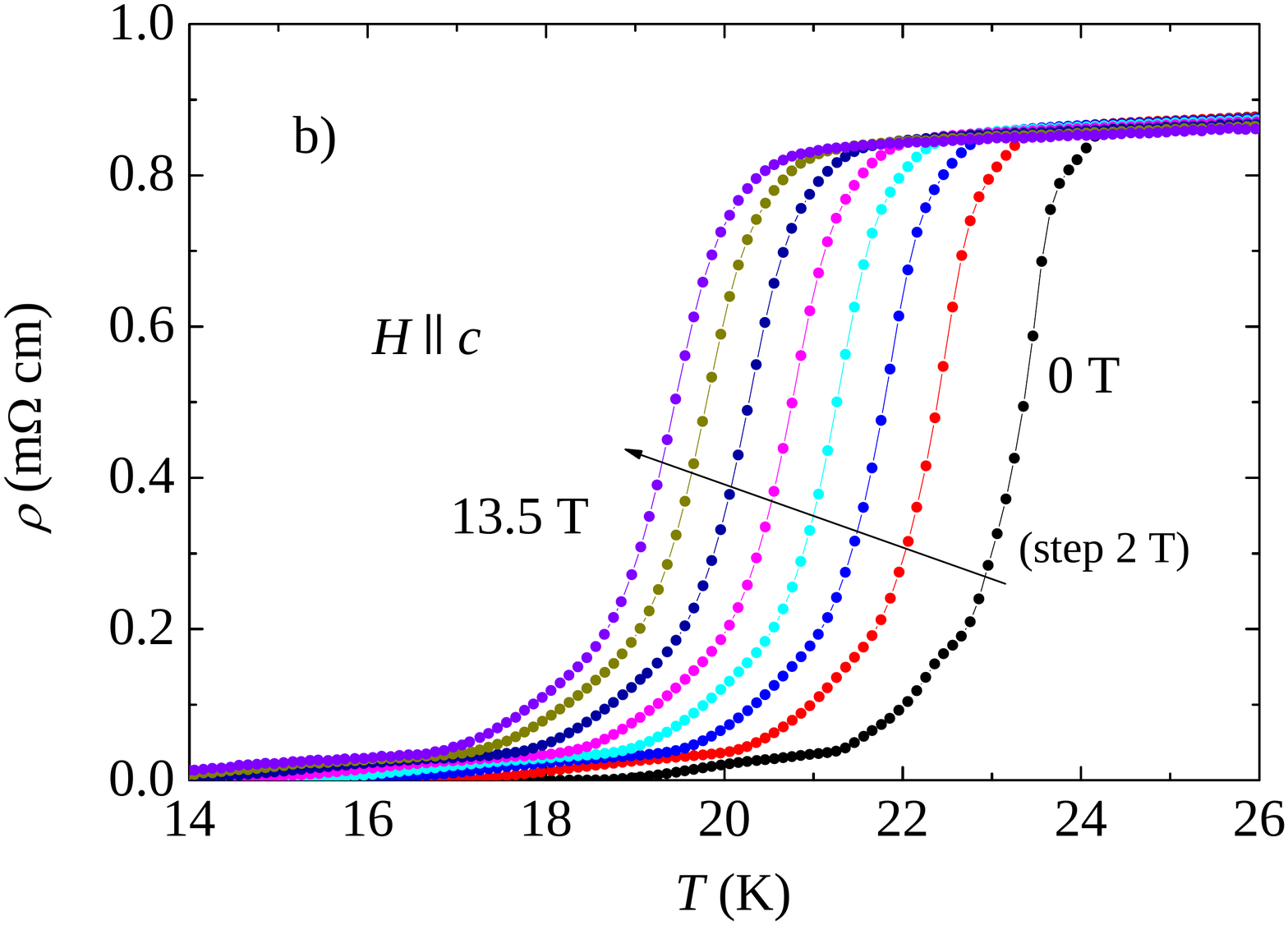}
\vspace{0cm}
\caption{(Color online) Examples of $\rho(T,H)$ dependences measured on the Ba$_{0.84}$Rb$_{0.10}$Sn$_{0.09}$Fe$_2$As$_{1.96}$ crystal with the field applied parallel to the (Fe$_2$As$_2$)-layers ($H||ab$) (a) and perpendicular to them ($H||c$) (b) in magnetic fields of 0, 2, 4, 6, 8, 10, 12, and 13.5 T. The solid lines are guides to the eye.}
\label{fig9}
\end{figure}

\subsection{Upper critical field and superconducting state anisotropy}

The electronic anisotropy is a determining factor for the behavior of a superconductor in an applied magnetic field, and thus is also of importance when possible applications are considered. We have estimated the anisotropy from the shift of the resistive transition as well as from magnetic torque measurements, leading to consistent results.

The resistance has been measured with the magnetic field applied parallel to the (Fe$_2$As$_2$)-layers ($H||ab$, $I||H$) and perpendicular to them ($H||c$, $I||ab$). Examples of $\rho(T,H)$ are shown in Fig. \ref{fig9}. Particularly notable is the well defined shift of the resistance drop with increasing field, without a significant broadening due to flux flow dissipation. In several of the larger crystals used for resistance measurements, the resistance drop proceeds in two or three steps, indicative of the presence of two or more distinct parts of the crystal with distinct transition temperatures. Yet, each of the steps can be followed as a function of field yielding very similar and consistent slopes of the upper critical field.

\begin{figure}[b!]
\vspace{0cm}
\includegraphics[width=1\linewidth]{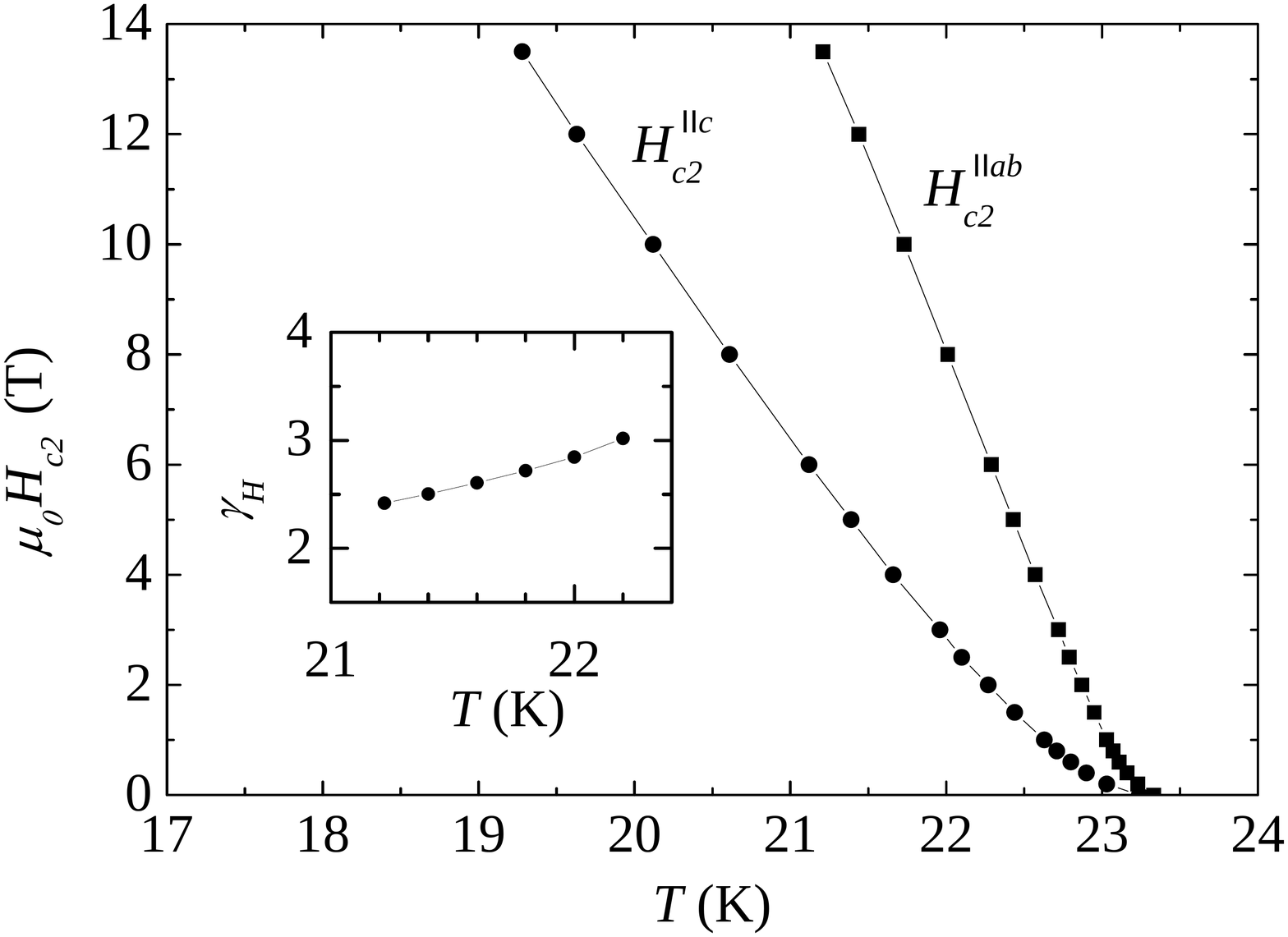}
\vspace{0cm}
\caption{Temperature dependence of the upper critical field with $H||ab$ and with $H||c$ for Ba$_{0.84}$Rb$_{0.10}$Sn$_{0.09}$Fe$_2$As$_{1.96}$. The solid lines are guides to the eye. Inset: The upper critical field anisotropy $\gamma_H=H_{\rm c2}^{||ab}/H_{\rm c2}^{||c}$ in the vicinity of $T_{\rm c}$. }
\label{fig10}
\end{figure}

As the resistive transitions are not markedly broadened, we define $T_{\rm c}(H)$ as the temperature where the resistance has decreased to 50 \% (see Fig. \ref{fig10}). For $H||ab$ the upper critical field $H_{\rm c2}^{||ab}$ increases with a slope of 7.1(3) T/K with a clearly visible rounding near $T_{\rm c}$, whereas $H_{\rm c2}^{||c}$ increases with a slope of 4.2(2) T/K again exhibiting an upturn near $T_{\rm c}$ where thermal phase fluctuations are expected to influence its determination. The upper critical slope $\mu_0dH_{\rm c2}^{||c}/dT$ is considerably higher than in YBa$_2$Cu$_3$O$_{7-\delta}$ (1.9 T/K),\cite{thirtyeight} in HgBa$_2$Ca$_2$Cu$_3$O$_{8+\delta}$ (2 T/K),\cite{thirtynine} in unsubstituted MgB$_2$ (0.12 T/K),\cite{forty} and in LaFeAsO$_{0.89}$F$_{0.11}$ (2 T/K).\cite{fortyone} This steep slope also points to a very high $H_{\rm c2}^{||c}(T=0)$. The upper critical field anisotropy in the vicinity of $T_{\rm c}$, defined as $\gamma_H=H_{\rm c2}^{||ab}/H_{\rm c2}^{||c}$, (inset to the Fig. \ref{fig10}) decreases with decreasing temperature, a similar trend as observed in the ``\emph{1111}'' pnictides.\cite{fortytwo} This suggests similar temperature dependences of the anisotropic parameters among different classes of pnictides.\cite{fortythree} However, it should be noted that the values for the upper critical field anisotropy are significantly smaller in the ``\emph{122}'' class ($\sim2.5-3$) than in the ``\emph{1111}'' group of superconductors ($\sim5-6$), with a $T_{\rm c}$ of $\sim45$ K.\cite{fortytwo}

A very important difference between the behavior of the ``\emph{122}'' and ``\emph{1111}'' iron-pnictide superconductors should be pointed out. While (Ba,Rb)122 shows a sharp transition and a clear onset of superconductivity, there is no sharp transition for ``\emph{1111}'' superconductors.\cite{fortytwo} Therefore, $T_{\rm c}(H)$ for ``\emph{1111}'' superconductors is less clearly defined and the determination of $H_{\rm c2}(T)$ and its anisotropy $\gamma_H(T)$ in the vicinity of $T_{\rm c}$ is criterion dependent. The results of high field resistivity measurements by Jaroszynski \emph{et al.}\cite{fortytwo} indicate that the upper critical field anisotropy becomes criterion independent for temperatures of about 10 K below $T_{\rm c}$ only, providing convincing evidence for the temperature dependent anisotropy in ``\emph{1111}'' superconductors for the limited temperature range of $\sim10-17$ K below $T_{\rm c}$. In contrast, the resistance drop at $T_{\rm c}$ in the (Ba,Rb)122 crystals remains essentially unaffected by high magnetic fields and, therefore, provides for a reliable determination of $H_{\rm c2}(T)$. Accordingly, the upper critical field anisotropy $\gamma_H(T)$ is well defined too (see Figs. \ref{fig9} and \ref{fig10}).

Estimating $H_{\rm c2}(0)$ from extrapolations of the present data, covering a limited temperature range, is inappropriate not only due to the curvature of $H_{\rm c2}(T)$, but more importantly because of the two-band nature of superconductivity. Nevertheless, if one would disregard these considerations, the simple Werthamer-Helfand-Hohenberg (WHH) extrapolation\cite{fortyfour} would give values of  $\mu_0H_{\rm c2}^{||c}(0)\simeq70(5)$ T and $\mu_0H_{\rm c2}^{||ab}(0)\simeq120(6)$ T.

\begin{figure}[t!]
\includegraphics[width=1\linewidth]{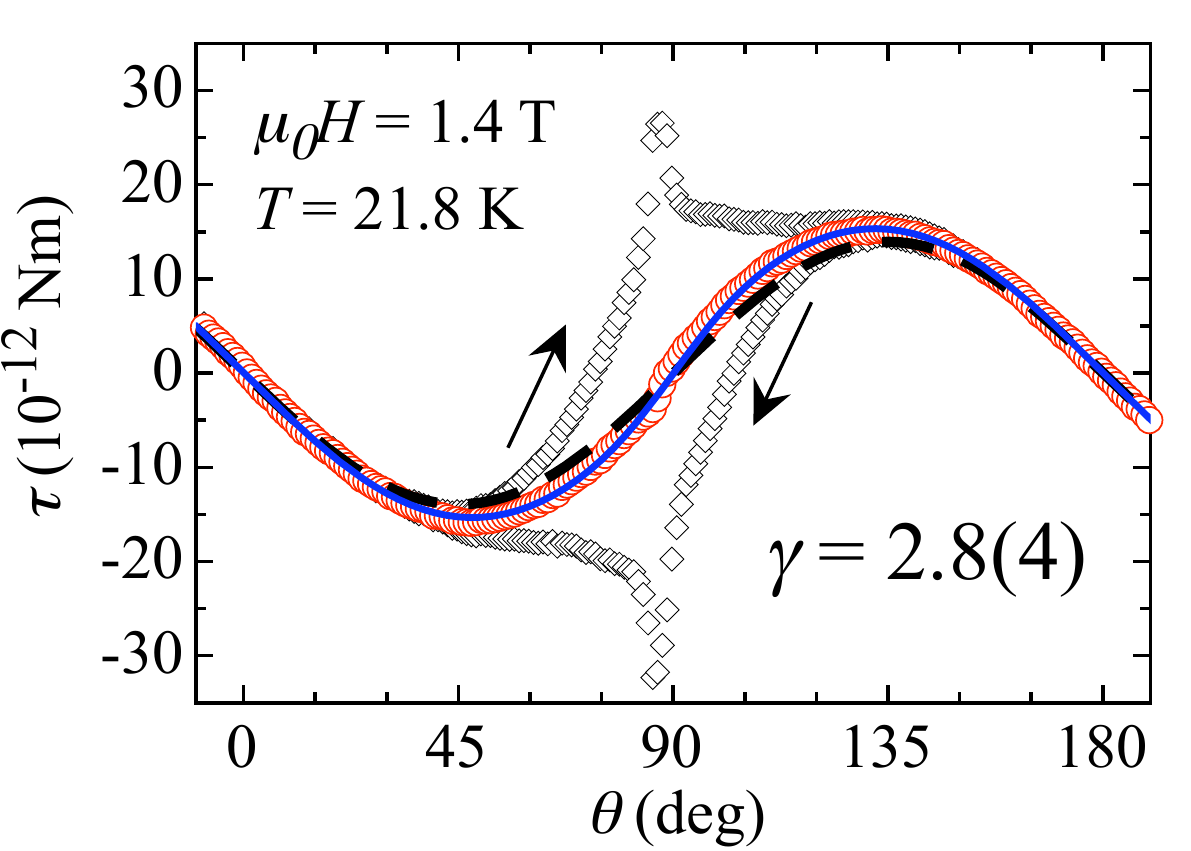}
\vspace{0cm}
\caption{(Color online) Angular dependence of the raw (black open diamonds) and the reversible (open circles) torque of the Ba$_{0.84}$Rb$_{0.10}$Sn$_{0.09}$Fe$_2$As$_{1.96}$ single crystal in the superconducting state at $T=21.8$ K and in $\mu_0H=1.4$ T. The solid line is given by the Kogan model (Eq. (\ref{kogan})) with the parameters fitted to the reversible torque data (open circles). A very big background component contributes a pronounced signal to the torque and is shown by dashed line.}
\label{fig11}
\end{figure}

Torque measurements were performed in the temperature range close to $T_{\rm c}$, where the pronounced irreversibility leads to only minor distortions of the torque. For minimizing pinning effects, the mean (reversible) torque, $\tau_{rev}=(\tau(\theta^+)+\tau(\theta^-))/2$ was calculated from measurements with counterclockwise and clockwise rotating of the magnetic field around the sample, as indicated by the black open diamonds in Fig. \ref{fig11}. Unfortunately, due to the small superconducting torque signal a big background component is contributing a pronounced additional signal to the torque. This sinusoidal component has been included by an additional sinusoidal function in the fitting model after Kogan \emph{et al.}\cite{fortyfive, fortysix}
\begin{equation}
\tau(\theta)=-\frac{V\Phi_0H}{16\pi\lambda_{ab}^2}\Bigg(1-\frac{1}{\gamma^2}\Bigg)\frac{\sin(2\theta)}{\epsilon(\theta)}\ln\Bigg(\frac{\eta H_{\rm c2}^{||c}}{\epsilon(\theta)H}\Bigg)+A\sin(2\theta),
\label{kogan}
\end{equation}
where $V$ is the volume of the crystal, $\Phi_0$ is the elementary flux quantum, $\gamma$ denotes the superconducting state anisotropy parameter, $\lambda_{ab}$ the in-plane component of the magnetic penetration depth, $H_{\rm c2}^{||c}$ the upper critical field along the $c$-axis of the crystal, $\eta$ denotes a numerical parameter of the order of unity depending on the structure of the flux-line lattice, $A$ is the amplitude of the background torque, and $\epsilon(\theta)=[\cos^2(\theta)+\gamma^{-2}\sin^2(\theta)]^{1/2}$.

The superconducting state anisotropy parameter $\gamma$, fitted to the mean torque data is found to be 2.8(4) near $T_{\rm c}$ in very good agreement with the estimate of $\gamma_H$ from resistivity measurements shown in Fig. \ref{fig10}. Due to the low anisotropy, the strong irreversibility and the pronounced normal state background, no temperature dependent study of the anisotropy parameter was performed. It is highly gratifying to find excellent agreement among the two ways for estimating the electronic anisotropy in the vicinity of $T_{\rm c}$, where independently of the predictions for the detailed electronic structure, all of the electronic anisotropies should coincide. Obviously, superconducting Ba$_{0.84}$Rb$_{0.10}$Sn$_{0.09}$Fe$_2$As$_{1.96}$ is much more isotropic than SmFeAsO$_{0.8}$F$_{0.2}$ and NdFeAsO$_{0.8}$F$_{0.2}$, where reliable anisotropy parameter values up to 20 can be derived.\cite{fortythree, twentyfive1} The anisotropy parameter measured for Ba$_{0.84}$Rb$_{0.10}$Sn$_{0.09}$Fe$_2$As$_{1.96}$ is much smaller than those typical for high-$T_{\rm c}$ superconductors, but it is quite similar to those reported for Ba$_{1-x}$K$_x$Fe$_2$As$_2$ and Sr$_{1-x}$K$_x$Fe$_2$As$_2$ \cite{twenty1, fortyseven, fortyeight}. The electronic coupling of the Fe$_2$As$_2$ layers through the intervening (Ba, Rb) layers is more effective than through the $Ln$O layers in the  ``\emph{1111}'' class of superconductors.

\subsection{Effect of Sn incorporation}

The effect of Sn incorporation on the properties of $A$Fe$_2$As$_2$ has not been studied in detail. Single crystals of BaFe$_2$As$_2$ grown from Sn flux reveal the structural and magnetic phase transition at significantly lower temperature (85 K)\cite{twenty1} than Sn-free material (140 K) although the magnetic structure is virtually the same.\cite{fortynine} The electrical resistivity (below the tetragonal-orthorombic transition) for BaFe$_2$As$_2$ with 1\% Sn increases upon decreasing temperature\cite{twenty1} while for pure BaFe$_2$As$_2$ crystals $\rho(T)$ is typical for metals.\cite{fifty} It is worth noting that although Sn incorporation lowers the structural transition temperature in BaFe$_2$As$_2$ it does not lead to superconductivity. The effect of Sn incorporation in superconducting alkali metal-substituted BaFe$_2$As$_2$ seems to be fully masked by the much stronger effect of the alkali metal doping.

\section{Summary}

Single crystals of Ba$_{1-x}$Rb$_x$Fe$_2$As$_2$ ($x$ = 0.05-0.1) have been grown and their crystallographic and basic superconducting state properties were presented. This is the first example of superconductivity induced by Rb substitution in this family of materials. It was found that the irreversibility line is located in relatively high magnetic fields comparable to the one for YBa$_2$Cu$_3$O$_{7-\delta}$ only, i.e. for the high-$T_{\rm c}$ superconductor with the lowest anisotropy. The electronic anisotropy, derived consistently from resistance and from magnetic torque measurements, ranges from $\sim$3 near $T_{\rm c}$ to lower values at lower temperatures, and indicates that Ba$_{1-x}$Rb$_x$Fe$_2$As$_2$ is electronically much more isotropic than SmFeAsO$_{0.8}$F$_{0.2}$ and NdFeAsO$_{0.8}$F$_{0.2}$. The critical current density at 5 K exceeds $10^6$ A/cm$^2$, which together with the high upper critical fields, is very promising for applications.

\section{Acknowledgments}
We would like to thank P. W\"agli for the EDX analysis. This work was supported by the Swiss National Science Foundation, by the NCCR program MaNEP, and partially by the Polish Ministry of Science and Higher Education within the research project for the years 2007-2009 (No.~N~N202~4132~33).

\end{document}